\documentclass[sigconf]{acmart}

\AtBeginDocument{%
  \providecommand\BibTeX{{%
    \normalfont B\kern-0.5em{\scshape i\kern-0.25em b}\kern-0.8em\TeX}}}

\setcopyright{none}
\renewcommand\footnotetextcopyrightpermission[1]{} 

 \acmConference[DI2KG '19]{Anchorage '19: 1ST INTERNATIONAL WORKSHOP ON
 CHALLENGES AND EXPERIENCES FROM DATA INTEGRATION TO KNOWLEDGE GRAPHS}{August 05, 2019}{Anchorage, Alaska, USA}

\begin{document}

\title{Tripartite Vector Representations for Better Job Recommendation}

\author{Mengshu Liu, Jingya Wang, Kareem Abdelfatah, Mohammed Korayem}
\affiliation{%
  \institution{CareerBuilder LLC}
  \streetaddress{5550 Peachtree Parkway}
  \city{Greater Atlanta Area}
  \state{Georgia, US}
  \postcode{43017-6221}
}
\email{{mengshu.liu, jingya.wang, kareem.abdelfatah, mohammed.korayem}@careerbuilder.com}




\renewcommand{\shortauthors}{Liu, et al.}

\begin{abstract}
Job recommendation is a crucial part of the online job recruitment business. To match the right person with the right job, a good representation of job postings is required. Such representations should ideally recommend jobs with fitting titles, aligned skill set, and reasonable commute. To address these aspects, we utilize three information graphs (job-job, skill-skill, job-skill) from historical job data to learn a joint representation for both job titles and skills in a shared latent space. This allows us to gain a representation of job postings/ resume  using both elements, which subsequently can be combined with location. In this paper, we first present how the representation of each component is obtained, and then we discuss how these different representations are combined together into one single space to acquire the final representation. 
The results of comparing the proposed methodology against different base-line methods show significant improvement in terms of relevancy.


\end{abstract}

\keywords{job recommendation, vector representation, location embedding}


\maketitle

\section{Introduction}
Online recruiting and job portals like Careerbuilder.com, Linkedin. com, and Indeed.com, have become the norm in the talent acquisition business. Millions of jobs are posted and even more resumes are uploaded daily. Different machine learning and information retrieval models have been applied to analyze these resumes and job descriptions, and multiple efforts have been made to match the two parties of the recruiting process. A good job/resume representation helps to improve many downstream products that in turn support the company\rq s goal of empowering employment and helping job seekers find jobs and the training they need. Specifically, it facilitates matching job seekers and employers by improving our search and recommendation products. \par

The content of a job posting incorporates all aspects of a position. Entities like job title, required skills, experience, degrees, benefits, company culture, location, etc, can be extracted and normalized. Among these, title, skills, and location are the top factors when defining a job position. While a job title determines the nature of a job, skills enrich the job title, and differentiate jobs with the same title by identifying their niceties. In order to offer a good job match, an accurate representation consisting of both is needed. Rather than simply using various word representations of job title and skills, recent works ~\cite{huang2015neural, tang2015user, mikolov2013distributed} start to consider both the text content and the relationship between matching pairs. In our previous work, we propose a novel representation learning based solution, which learns job and skill vector representations into a shared latent space using three pre-processed graphs ~\cite{dave2018combined}. To extend this work, we consider the interconnection of both job title and skills to learn the vector representation of job postings/ resumes. By using a retrofitting model similar to the work of ~\cite{faruqui15}, we combine the pre-trained representations of job title and skills into one vector to represent a job or resume. In terms of location, majority of job seekers prefer jobs within reasonable commute distance. For many people, relocation is not an option, and short commute is always a big plus. Similarly, for most companies, remote employees are not preferred either. Therefore, we explicitly include a location vector in our representation. To achieve quick, accurate vector search and recommendation, we utilize FAISS ~\cite{JDH17}, which is a library for efficient similarity search and clustering of dense vectors.

Our contributions in this paper are as follows:

$\bullet$ The vector representation proposed is applicable for both job postings and resumes. It's not only a flexible representation to obtain similar jobs or similar candidates, but also provides a direct mapping of jobs and resumes.\par
$\bullet$ We generate a more holistic vector representation jointly learned for both titles and skills which can be used in the job recommendation system.\par
$\bullet$ We incorporate the location explicitly included in our representation.\par
$\bullet$ We employ retrofitting to refine the job and skill vectors using semantic lexicons detailed in Section ~\ref{retrofitting}.
\begin{figure}[htp]
\centering
\includegraphics[width=8.5cm]{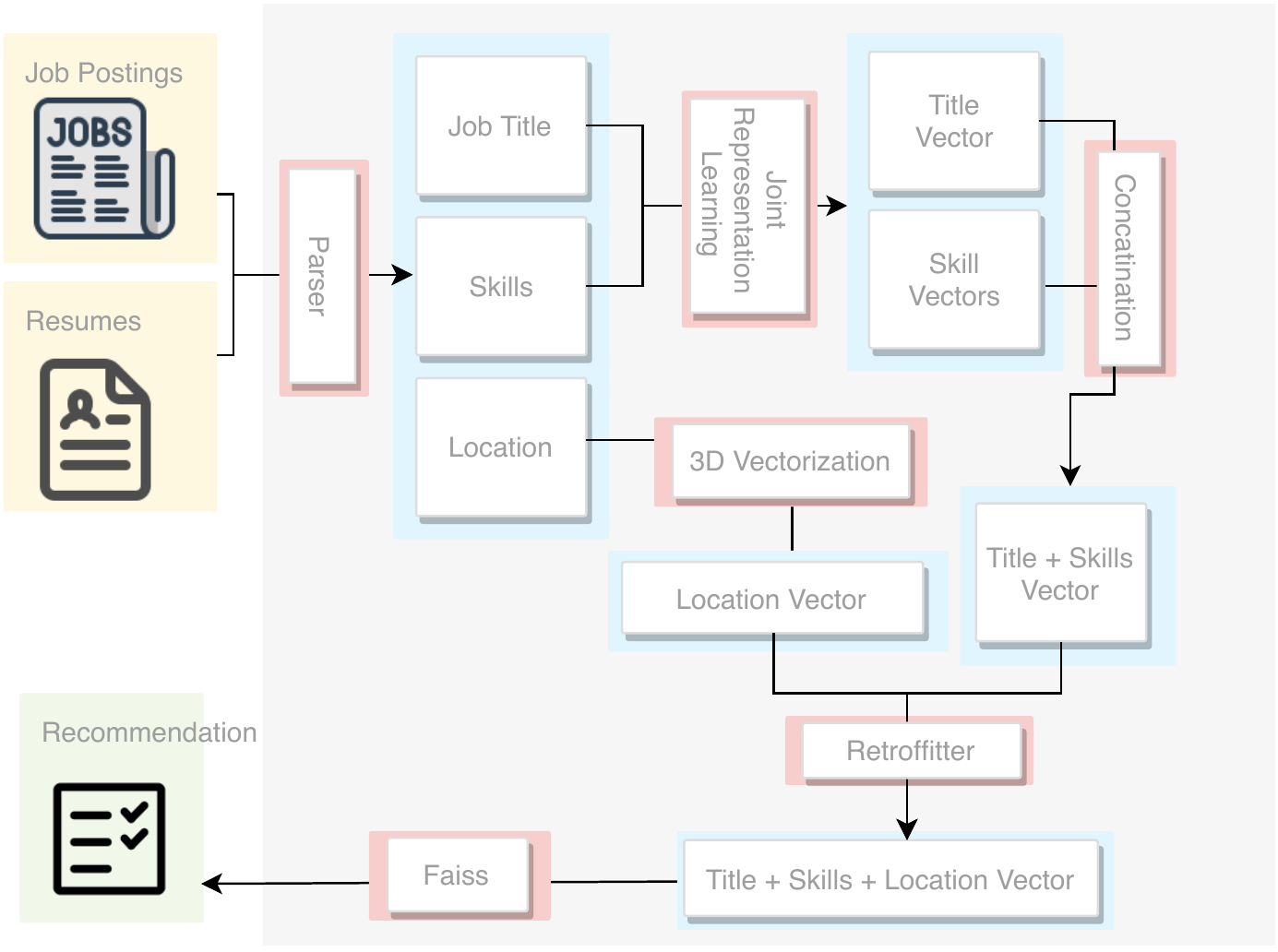}
\caption{Workflow to generate recommendations using our tripartite vector representations.}
\label{fig:flowchart}
\end{figure}

\section{Related Work}
The two major solutions of job recommendation in existing works are content-based representation learning and collaborative filtering (CF).
CF builds a very high dimension sparse matrix to maintain the relationship between user and product ~\cite{sedhain2015autorec, wang2015collaborative}. In our paper, this relationship is between each pair of job and skill. Inspired by the natural language Processing approach (NLP) of skip-gram~\cite{mikolov2013distributed}, ~\cite{barkan2016item2vec} proposes to learn item embedding based on interaction history in the form of a sequence. ~\cite{wu2014browsemaps} presents an architecture of CF and points out the common problem of cold start. CF relies on historical interactions, therefore, when a new candidate registers or a new job is posted, no interaction record is available. Content-based method on the other hand, makes recommendations according to the characteristics of users and products as in ~\cite{van2013deep}.

Most recent works, including our proposed system, use hybrid approaches of both profile properties and interaction history ~\cite{oh2014personalized}. ~\cite{huang2015neural, tang2015user} jointly learns feature representation of users and their selected items in one convolutional neural network. ~\cite{covington2016deep} concatenates the learned sparse and dense representation of user's activity history, query, and profile as input. In this case, the recommendation problem is treated as an extreme multi-class classification problem. The classifier is a series of non-linear activation functions followed by a softmax. ~\cite{li2017nemo} employs a convolutional neural network to encode the user profile followed by a Long short-term memory (LSTM) ~\cite{sak2014long} to encode the user's interaction history in training and continue the LSTM on time axis as a decoder to predict user's next activity in testing.

Our work is similar to the hybrid approaches in terms of learning the embedding of job titles and skills based on both their characteristics and interactions. Meanwhile, it is different that, instead of simply pairing the related jobs and skills, our profile representation models three specific relationships simultaneously between jobs and required skills. Moreover, we employ retrofitting to achieve better representation.

\section{Methodology}
In this section, we describe the design and methodology of our tripartite vector representation. A job posting contains important information of the hiring position. Most importantly it includes job title, skills, and location. Here, we use a combined representation vectors of job titles and skills, and obtain a vector representation for location transformed from latitude and longitude. Figure ~\ref{fig:flowchart} shows the design of our method. Job posting and resume data go to an in-house job parser, where job title/skills/location are extracted. While titles and skills are jointly trained by a representation learning framework (Section ~\ref{ts}. Location information is also vectorized and normalized (Section ~\ref{lv}). In order to combine the tripartite, the title representation is first retrofitted by a list of skills associated with each job before added with the location vector (Section ~\ref{retrofitting}).

\subsection{Job Title and Skill Vector} \label{ts}
Title and skills are the defining features of any job. Embeddings for both titles and skills are learned in the same k-dimensional space ~\cite{dave2018combined}, by utilizing three types of information networks from historical job data: (i) job-job transition network, (ii) skill-skill co-occurrence network, and (iii) job-skill co-occurrence network. Our goal is to encode the local neighborhood structures captured by the three networks.

For the job-job transition graph, we assume that the transition between similar jobs $x$ and $y$ is more likely to happen than non-similar jobs $x$ and $z$. Let $A_{xy}^j=\langle w_x, w_y\rangle$, the dot product of the two embedding vectors, be the affinity score between job $x$ and job $y$, and $D_{jj}$ (job-job) represents the transition relationship of job triplets $(x, y, z)$. The objective is to learn representation $W$ so that
\begin{equation}
    O^{jj} = \min_{W}-\sum_{(x,y,z)\in D^{jj}} \ln{\sigma (A_{xy}^{jj}-A_{xz}^{jj})}
\end{equation}
Where sigmoid function $\sigma(\nu) = \frac{1}{1+e^{-\nu}}$ is used as the probability function which preserves the order $A_{xy}^j>A_{xz}^j$.\par
Similarly, for the skill-skill graph, coexisting skills $x$ and $y$ which appear on the same job posting or the same resume are closer to each other than non-coexisting skills such as $x$ and $z$. Let $D^{ss}$ (skill-skill) be the set of training triplets of skills with coexisting relationship, our objective here is
\begin{equation}
    O^{ss} = \min_{W'}-\sum_{(x,y,z)\in D^{ss}} \ln{\sigma (A_{xy}^{ss}-A_{xz}^{ss})}
\end{equation}
Moreover, for the job-skill graph, if skill $y^s$ appears on the advertisement of job $x^j$, its embedding vector is more similar to $x^j$ than non-related skill $z^s$. Given the set of training triplets $D^{js}$ (job-skill), our desired vector representation of jobs $W$ and skills $W'$ are learned according to the objective 
\begin{equation}
    O^{js}=\min_{WW'}-\sum_{(x^j,y^s,z^s)\in D^{js}} \ln{\sigma (A_{xy}^{js}-A_{xz}^{js})}
\end{equation}

Finally, to achieve high quality job and skill embedding, we optimize this joint objective function 
\begin{equation}
    O(W,W')=\min_{W,W'}O^{jj}+O^{ss}+O^{js}+\lambda \cdot{(\mid\mid W\mid\mid_F^2 +\mid\mid W'\mid\mid_F^2)}
\end{equation}
where $\lambda$ is the coefficient of the $l_2$ regularization term $\mid\mid \cdot \mid\mid_F^2$ to avoid over-fitting. \par
Vectors of dimension size 50 are obtained for 4325 unique job titles and 6214 skills, using joint Bayesian Personalized Ranking (BPR) ~\cite{Rendle:2009:BBP:1795114.1795167}.

\subsection{Location Vector} \label{lv}
Location is another key factor in a job posting. Commute time matters when people are choosing a position. The common practice when dealing with location is to pre-filter or post-filter the recommendations with a fixed radius. This method has a few downsides: 1. Some jobs are less sensitive to distance than others. For example, people are more willing to commute longer with a highly compensated job than a minimum wage part-time one; 2. The system is more difficult to implement because of the extra layer of filtering. 3. Specifically for our case where Faiss is used for similarity search, only one index file is needed if location is embedded in the vector. \par
\begin{figure}[htp]
\centering
\includegraphics[width=8cm]{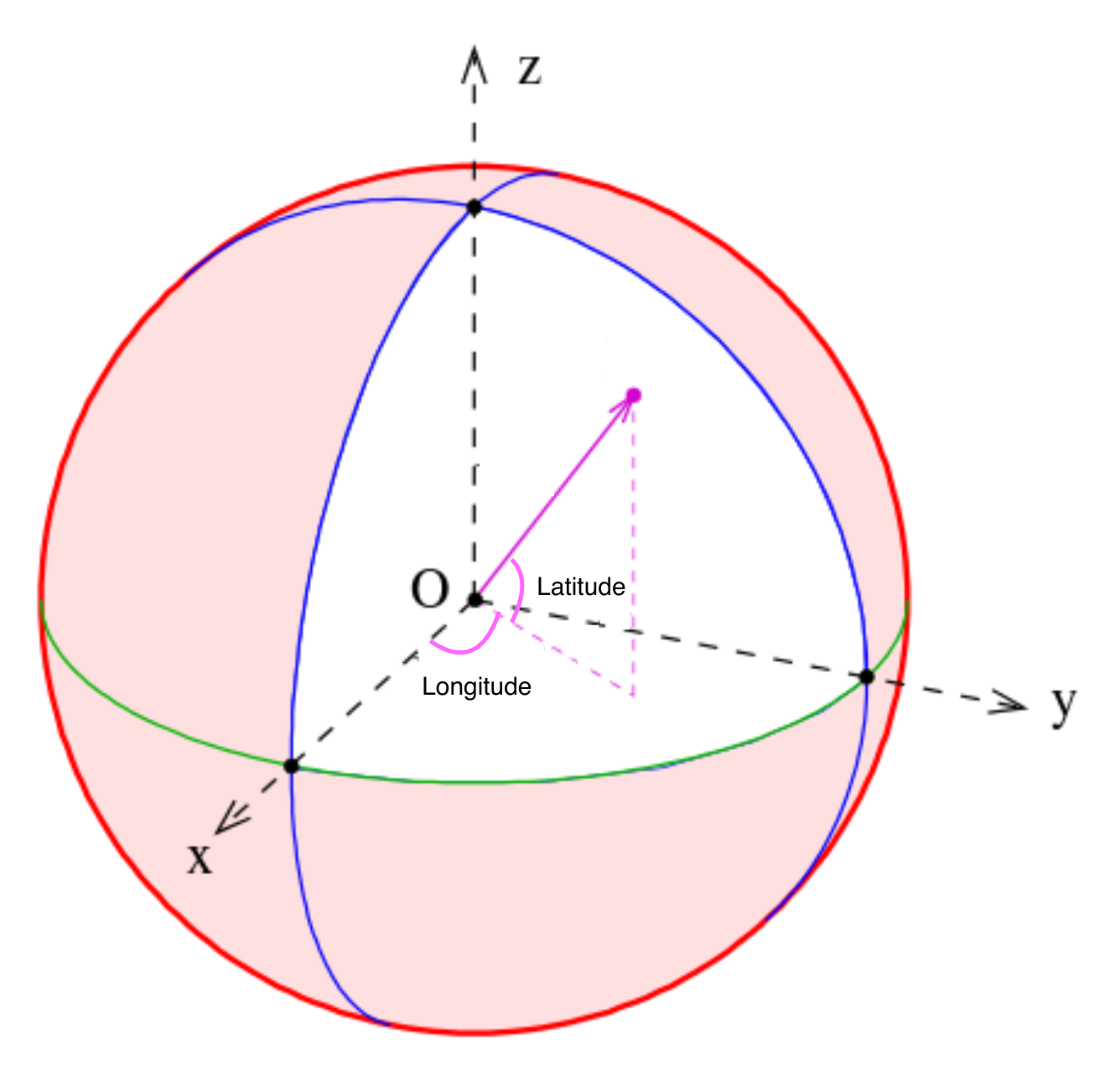}
\caption{Latitude and longitude can be converted into a three-dimensional coordinates.}
\label{fig:Latitude and longitude conversion}
\end{figure}

Before latitude and longitude can be added to our embedding model, they need to be transformed since they are not on the same scale as the title and skill vector. Geo-locations are three dimensional in its nature, and to represent location in a similar fashion as title and skill vectors, we did a transformation of latitude and longitude, as shown in Figure~\ref{fig:Latitude and longitude conversion}:


\begin{equation}
x = \cos(\theta) \times \cos(\phi) 
\end{equation}

\begin{equation}
y = \cos(\theta) \times \sin(\phi) 
\end{equation}

\begin{equation}
z = \sin(\theta)
\end{equation}

Where $\theta$ represents the latitude and $\phi$ is the longitude. 

Thus, location is represented as a normalized three dimensional vector, which can be later combined with the title and skill vector.

\subsection{Combination of title, skill, location}\label{retrofitting}

Job postings with the same title might require different skill sets at different companies in different industries. The next natural question is how to combine the title and skill vectors together to a personalized vector for a specific job, given both title and skill vectors are trained in the same latent space.\par
To assemble the vectors back into one to represent the job posting, we apply the retrofitting method ~\cite{faruqui15} to combine the vectors of the job title and all skills of the job posting. This method adjusts the position of the job title based on the skills appeared in the job posting. For example, consider one job posting looking for a web developer, and a person has a recent title of JavaScript developer. The two job titles are similar, though still different. If JavaScript is listed as a top skill in the job posting, the distance between the job and the resume will be shortened. This gives the person a better place in ranking, even though the job title is not a perfect match.

In Faruqui's work ~\cite{faruqui15}, they implemented a retrofitting method to adjust any pre-trained embedding using semantic lexicons:
\begin{equation}
\boldsymbol{q}_i = \frac{\sum_{j:(i,j)}\beta_{\text{ij}}\boldsymbol{q}_j+\alpha \hat{\boldsymbol{q}_i}}{\sum_{j:(i,j)}\beta_{\text{ij}}+\alpha_i}
\end{equation}
where $\boldsymbol{q}_i$ is the modified vector, $\hat{\boldsymbol{q}_i}$ is the initial vector, $\boldsymbol{q}_j$ are the neighbors.
This is derived by minimizing the distance between initial $q_i$ and neighbors. And for their case, the results converge after 10 iterations.\par

\begin{figure}[htp]
\centering
\includegraphics[width=8.5cm]{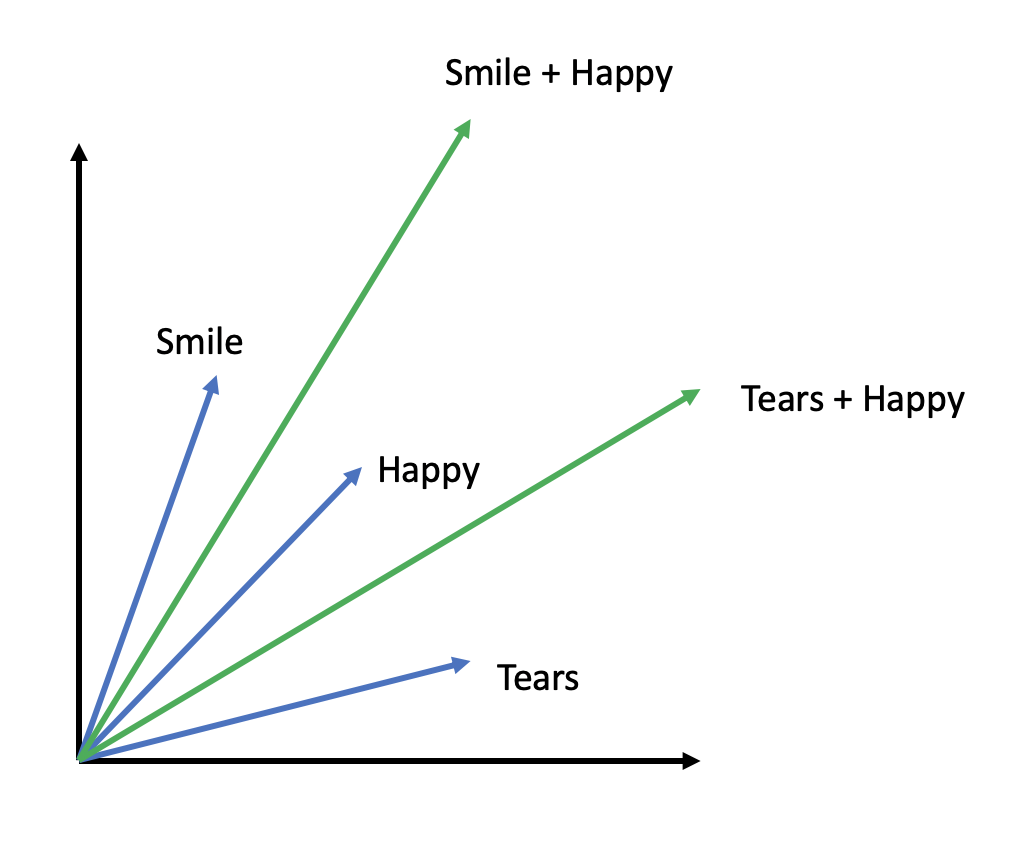}
\caption{Addition of embedding vectors.}
\label{fig:vector}
\end{figure}

Intuitively, if we add two vectors together, the result vector will lie in between the initial two vectors. Adjusting vectors by adding the vectors of neighboring words will bring words similar in meanings closer to each other. For example, in Figure~\ref{fig:vector}, $\hat{\boldsymbol{Smile}}$ and $\hat{\boldsymbol{Tears}}$ are the initial vectors for two very different words, so they are further apart in direction (cosine similarity). After the tuning using a neighboring word: $\boldsymbol{happy}$, the new vectors for smile and tears are closer than before, since the tears are in the context of "happy tears".\par

Similarly, we applied the above operation to our job posting data, and generated a combined representation of the job posting using title and the skills. Jobs with similar set of skills as well as the same title will have a higher rank than the ones with the same title but less overlapping skills. Rather than going through several iterations as in the original paper, we only apply the updates once, as skill vectors are not updated in the process. We can view this one time update as fine tuning of the title vector, or the calculation of the "job + title" vector, shown as below, where n is the number of skills in a job posting, which are given the same weight:
\begin{equation}
\boldsymbol{q}_\text{job} = \frac{n\cdot{\boldsymbol{q}_\text{title}}+\sum{\boldsymbol{q}_\text{skill}}}{2n}
\end{equation}

Location vector is then concatenated with the new vector. The weight of the location vector can be adjusted based on how the jobs are sensitive to distance.

\begin{equation}
\boldsymbol{q}_\text{job+location} = [\boldsymbol{q}_\text{job},x,y,z]
\end{equation}

\section{Results}
To test and evaluate our proposed system, we use a real dataset (jobs and users) via CareerBuilder.com. CareerBuilder operates the largest job posting board in the U.S. and has an extensive growing global presence, with millions of job postings, more than 60 million actively searchable resumes, over one billion searchable documents, and more than a million searches per hour. \par
Each job/ resume is parsed into job title, skill list, and location (Figure ~\ref{fig:flowchart}). We then translate these three parts into their vector representations, which are later combined to the final job posting vector. We have also curated a list of top skills for each job title, in case skills are not available. The curation is based on millions of resumes and job postings, as well as human knowledge. We use this skill list to generate a vector if no skill is given. This provides a more accurate representation of the job than just using a job title. These vectors all go into our recommendation engine for different recommendation products.  \par
We calculate the vectors on 300K actual job postings on the Careerbuilder Website, using 6 different models (FastText, Word2Vec, Glove-6B-300 , Glove-840B-300 ~\cite{bojanowski2017enriching, NIPS2013_5021, pennington2014glove}, and our proposed retroffiter model with and without location embedding), so each job posting has 6 vectors for comparison. For each of these vectors, we calculate the top 50 similar vectors using FAISS. For job posting data, a flat index method is used. For resume data, which is much larger in size, we compress them to 16 blocks in 64-dimension and build the inverted indexing of size 65535. Four metrics are computed for evaluation: 1) Distance, which is the geographic distance between the recommended job and the input. 2) In-range job counts, which is the number of jobs within 50 miles in the top 50 recommended jobs. 3) Title match rate, which is the percentage of recommended jobs with the same job title as the input. 4) Title coverage, which measures the title match rate within 50 miles. The results for different models are computed and compared. \par
Table 1 shows that by explicitly including latitude and longitude in our embedding, the average distance of recommended jobs is reduced by 90\%. All top 50 jobs are within a reasonable proximity of the original input job.  \par
Table 2 shows the number of jobs within 50 miles in the top 50 jobs increased dramatically for retrofitter with location embedding. Both table 1 and table 2 show that the base-line methods are not location sensitive, even though location is included in the text file. Implicitly having location in the embedding is far from enough, and the advantage of explicitly including location in the embedding model is prevailing. \par
Table 3 shows title match rate and coverage. Our retrofitter model with no location embedding has a much higher title match rate and coverage than the base-line methods as well as the retrofitter model with location. However, the drop in performance in our location retrofitter model purely comes from the limited relevant jobs in close proximity, which is a trade-off we have to make. A perfect job match from ten thousand miles away is simply not a perfect job for most people. Even so, our retrofitter model still has a better results in both measurements than the base-line methods. \par

\begin{table}[]
\begin{tabular}{|l|l|l|l|l|}
\hline
Model&  Average& Median & STD \\ \hline
 FastText& 940.85 & 760.69 & 819.33 \\ \hline
 W2V-300&  936.46 & 754.67 & 818.58 \\ \hline
 Glove-6B-300&  959.54 &785.40  & 823.14 \\ \hline
 Glove-840B-300&  942.22 &763.72 & 816.83 \\ \hline
 Retrofitter-no loc&  902.918 & 727.087 & 836.275 \\ \hline
 Retrofitter-loc&  \textbf{90.417} & \textbf{70.177} & \textbf{95.673} \\ \hline
\end{tabular}
\caption{Distance statistics for different representations.}
\end{table}

\begin{table}[]
\begin{tabular}{|l|l|l|l|l|}
\hline
Model &  Average& Median \\ \hline
 FastText-300&  5.637 & 3.0  \\ \hline
 W2V-300&  5.67 & 3.0  \\ \hline
 Glove-6B-300& 5.108 &3.0   \\ \hline
 Glove-840B-300&  5.503 &3.0 \\ \hline
 Retrofitter-no loc& 1.614 &1.0 \\ \hline
 Retrofitter-loc& \textbf{21.420} & \textbf{20.0} \\ \hline
\end{tabular}
\caption{Number of jobs with a distance less than 50 miles in the top 50 jobs.}
\end{table}

\begin{table}[]
\begin{tabular}{|l|l|l|l|l|}
\hline
Model &  Carotene-Match (\%) & Coverage (\%) \\ \hline
 FastText-300&  11.0 & 3.1  \\ \hline
 W2V-300&  11.2 & 3.1  \\ \hline
 Glove-6B-300& 10.6 &3.0   \\ \hline
 Glove-840B-300&  10.6 &3.0 \\ \hline
 Retrofitter-no loc&  \textbf{68.9} & \textbf{14.7} \\ \hline
 Retrofitter-loc& \textbf{14.1} & \textbf{5.5} \\ \hline
\end{tabular}
\caption{This table shows two metrics. First, it shows the percentage of jobs with the same title match. Second, it shows the coverage which means the percentage of recommended jobs within the 50 mi with the same job title. }
\end{table}
\section{Conclusion and Future Work}
In this paper, we discuss the representation model which can be used for a  recommendation system and it is currently being utilized within CareerBuilder. Three facets of a job posting are considered: job title, job skills, and location. While job title carries the most weight in determining what a job is, skill set defines the nuances which differs from job to job. Most job seekers are also very sensitive to the location of a job. In our model, we encompass all three of these aspects, and are able to give location sensitive, highly related job recommendations to our users. \par
There are a number of improvements being worked on such as : Develop an inductive learning framework to accommodate newly emerged job titles and skills, as the current model is transductive, and representation vectors only exist if it is in the input graph; Incorporate more features in the job representation such as education and previous experience; Adjust the location embedding in a more quantifiable way to control the radius of recommended jobs; Combine the current representation model with other models to provide a better results; Apply different weights to skills based on their importance to the job title.

\newpage
\bibliographystyle{unsrt}
\bibliography{ref}
\end{document}